\pdfoutput=1

\documentclass[sn-mathphys-ay]{sn-jnl}

\usepackage{graphicx}
\usepackage[table,xcdraw]{xcolor}
\usepackage{multirow}
\usepackage{amsmath,amssymb,amsfonts}
\usepackage{amsthm}
\usepackage{mathrsfs}
\usepackage[title]{appendix}
\usepackage{textcomp}
\usepackage{manyfoot}
\usepackage{booktabs}
\usepackage{algorithm}
\usepackage{algpseudocode}
\usepackage{listings}
\usepackage{CJKutf8}
\usepackage{mdframed}
\usepackage{tabularx}
\usepackage{caption}
\usepackage{hyperref} 

\theoremstyle{thmstyleone}

\theoremstyle{thmstyletwo}%

\theoremstyle{thmstylethree}%

\begin{document}

\title[Article Title]{Adaptive Two-Phase Finetuning LLMs for Japanese Legal Text Retrieval}

\author*[1,]{\fnm{Quang Hoang} \sur {Trung}}\email{trung.quang@vj-tech.jp}

\author[1]{\fnm{Nguyen Van Hoang} \sur{Phuc}}

\author[1]{\fnm{Le Trung} \sur{Hoang}}

\author[1,2]{\fnm{Quang} \sur{Huu Hieu}} \email{hieuquang@aj-tech.jp}

\author[3,4,1]{\fnm{Vo} \sur{Nguyen Le Duy}}\email{duyvnl@vj-tech.jp, duyvnl@uit.edu.vn}

\affil[1]{\orgname{VJ Technologies}, \city{Da Nang City}, \country{Vietnam}}

\affil[2]{\orgname{AJ Technologies}, \city{Nagoya City}, \country{Japan}}

\affil[3]{\orgname{University of Information Technology}, \city{Ho Chi Minh City}, \country{Vietnam}}

\affil[4]{\orgname{Vietnam National University}, \city{Ho Chi Minh City}, \country{Vietnam}}

\abstract{
Text Retrieval (TR) involves finding and retrieving text-based content relevant to a user’s query from a large repository, with applications in real-world scenarios such as legal document retrieval. While most existing studies focus on English, limited work addresses Japanese contexts. In this paper, we introduce a new dataset specifically designed for Japanese legal contexts and propose a novel two-phase pipeline tailored to this domain. In the first phase, the model learns a broad understanding of global contexts, enhancing its generalization and adaptability to diverse queries. In the second phase, the model is fine-tuned to address complex queries specific to legal scenarios. Extensive experiments are conducted to demonstrate the superior performance of our method, which outperforms existing baselines. Furthermore, our pipeline proves effective in English contexts, surpassing comparable baselines on the MS MARCO dataset. We have made our code publicly available on GitHub, and the model checkpoints are accessible via HuggingFace.
}

\keywords{Text Retrieval, Large Language Models, Legal Document Retrieval, Japanese Legal Dataset, Ensemble Model}

\maketitle

\section{Introduction}
Text Retrieval (TR) is the task of finding and retrieving text-based content relevant to a user’s query from a large repository. TR has significantly evolved from simple term-matching models \citep{salton1983modern} to more advanced methods that capture semantic nuances. Initially, TR utilized Boolean and vector space models \citep{salton1975vector}, focusing on keyword matching and measuring lexical similarity between user queries and documents. The development of inverted indexes and statistical language models enhanced efficiency and incorporated contextual understanding into retrieval tasks \citep{song1999general,martineau2009delta}. 

TR is fundamental to systems such as search engines, dialogue platforms, and question-answering services, making it vital in both personal and professional contexts. It is widely applied in web search \citep{bajaj2016ms}, open-domain question answering \citep{chen-etal-2017-reading}, and fact verification \citep{thorne-etal-2018-fever}. 
Additionally, TR enhances the performance of large language models (LLMs) in retrieval-augmented generation (RAG) pipelines \citep{lewis2020retrieval, shi2023replug}. Recent advancements with LLMs, such as LLaMA \citep{touvron2023a, touvron2023b}, GPT-4 \citep{achiam2023gpt}, have greatly improved TR by facilitating better query interpretation and more contextually rich responses. Foundational contributions, including BM25 \citep{robertson1995okapi}, BERT \citep{devlin-etal-2019-bert} and T5 \citep{raffel2020exploring}, have shaped TR’s development, establishing its indispensability for applications ranging from simple web searches to complex, data-driven decision-making.
\begin{figure}[!t]
    \centering
    \includegraphics[width=0.9\textwidth]{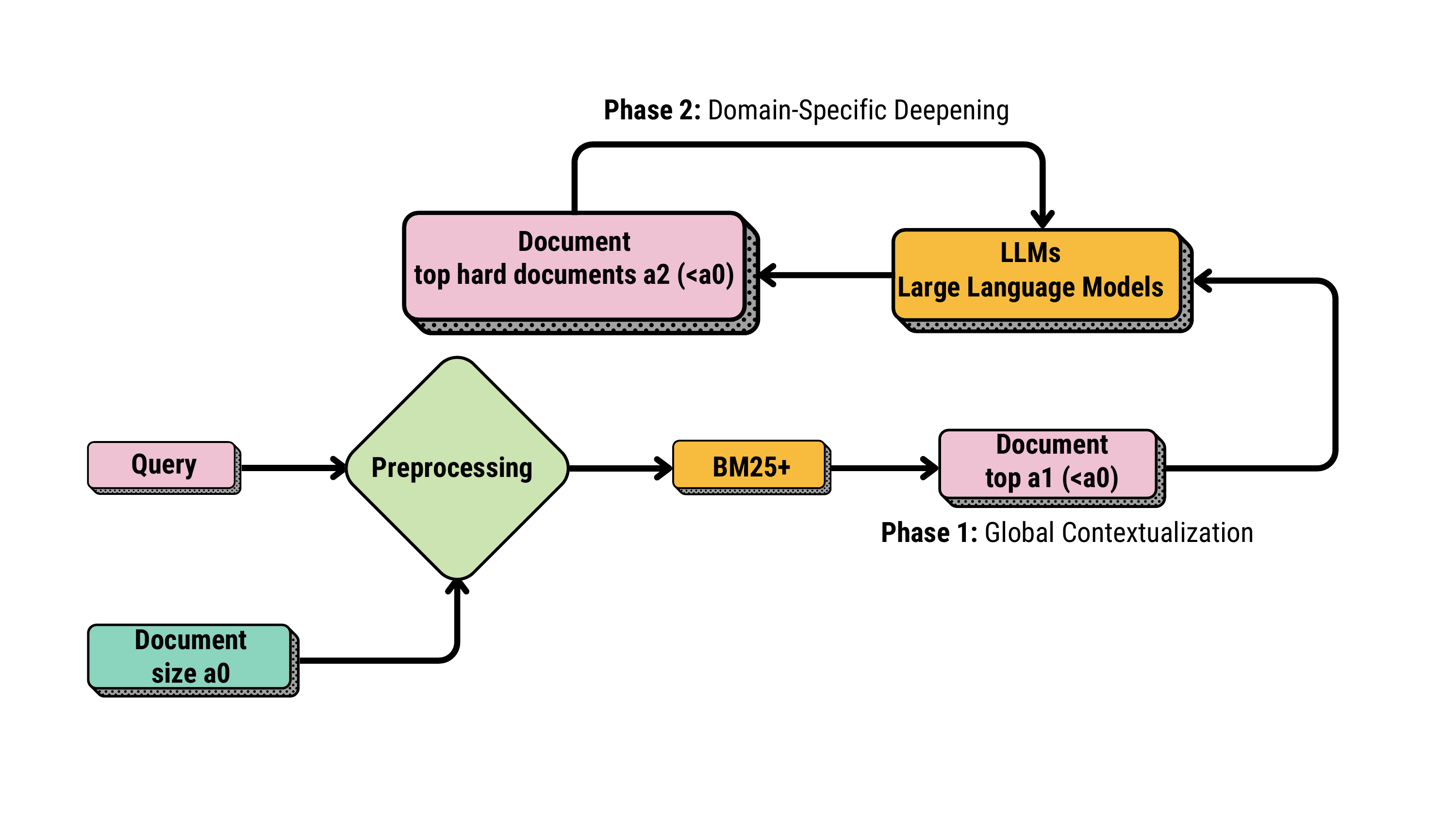}
    \caption{The overview of the multi-stage retrieval applies a two-phase approach for optimal performance. During the Preprocessing stage, unnecessary elements such as special characters, undesired symbols, and stop words are filtered out, ensuring the input text is cleaner and more accurate before passing through BM25+. This process allows for the retrieval of top-$a_1$ relevant documents related to the query. Starting with \textbf{Phase 1}: Global Contextualization, the model leverages three types of documents: positive (human-labeled documents that are truly relevant to the query), negative (top documents retrieved by BM25+ that appear relevant to the query but are not genuinely positive), and easy (top documents relevant to other queries, which may be irrelevant to the current query, incorporated via in-batch negatives). This diverse input enriches the training process, enabling the model to develop a broad global understanding and enhance generalization. In \textbf{Phase 2:} Domain-Specific Deepening, after the model has been fine-tuned in \textbf{Phase 1}, it continues to fine-tune on top-$a_2$ hard documents specific to the query. These include positive and hard negatives documents (highly relevant documents retrieved by the \textbf{Phase 1} fine-tuned model but not labeled as positive). This targeted refinement enables the model to focus on documents that are difficult to distinguish as relevant or not to the query, significantly enhancing precision and performance in complex retrieval scenarios.}
    \label{fig:image_1}
\end{figure}

The field of TR has made substantial progress with various methods for English. However, there is limited advancement for Japanese retrieval, posing challenges when adapting English-based techniques. A few methods have proven effective for English and are now being applied to Japanese, though they encounter significant limitations due to linguistic differences. Specifically, we explore the limitations of sparse retrieval, dense retrieval, and generative retrieval on Japanese datasets.

Sparse retrieval, which relies on term-matching methods like TF-IDF \citep{SALTON1988513}, BM25 \citep{robertson2004simple}, BM25+ \citep{robertson1994some}, struggles with Japanese due to its complex morphology, flexible syntax, and the nuances of word segmentation, which complicate accurate term matching. 

Dense retrieval, which relies on embeddings for semantic similarity, has potential in Japanese but requires large, specialized datasets, a resource often lacking. The integration of large language models (LLMs) has propelled dense retrieval by improving both in-domain accuracy and cross-domain generalization, as well as enabling new capabilities like instruction-following and in-context learning. Early advances with models like BERT \citep{devlin-etal-2019-bert} and T5 \citep{raffel2020exploring} laid the groundwork, but recent LLMs such as LLaMA \citep{touvron2023a, touvron2023b} and RepLLaMA \citep{ma2024fine} have set new benchmarks, outperforming traditional multi-stage methods on tasks like MS-MARCO \citep{bajaj2016ms} and BEIR \citep{thakur2021beir}. Efforts such as Llama2Vec \citep{li2023making} and NV-Embed \citep{lee2024nv} have further optimized these models, enhancing adaptability and performance. LLM-based dense retrieval models now offer advanced functionalities, handling diverse queries with better adaptability, flexibility, and efficiency than earlier methods. 

Generative retrieval, where models directly generate document identifiers (DocIDs) or content, bypasses traditional index systems by embedding corpus knowledge within model parameters, simplifying retrieval for open-ended tasks. However, it faces computational challenges and requires high-quality Japanese data, which is often limited. Techniques like the DSI model \citep{tay2022transformer}, which fine-tunes LLMs (e.g., T5) for DocID generation, and prompting-based methods using models like GPT for URL generation, offer promising results but require large models and extensive datasets. Key challenges remain, including balancing efficiency, response times, and domain-specific accuracy through fine tuning. These limitations underscore the need for language-specific adaptations to address Japanese retrieval requirements effectively, highlighting the challenges of directly transferring English based methodologies.

In this paper, we present a novel dataset tailored specifically for the Japanese legal domain and propose a specialized retrieval pipeline tailored to the unique demands of Japanese legal text retrieval, consisting of two primary phases as shown in Figure \ref{fig:image_1}. In \textbf{Phase 1}, the model learns a broad understanding of global contexts by leveraging positive, negative, and easy documents. The positive documents are human-labeled documents that are definitively relevant to the query. In contrast, the negative documents consist of documents that appear highly relevant to the query (e.g., top-ranked documents retrieved via BM25+) but are, in reality, not relevant to it. Meanwhile, the easy documents are derived from the top-$a_1$ documents relevant to other queries and are generally less relevant to the query being evaluated. These easy documents, integrated through the in-batch negative mechanism during training, diversify the dataset and expose the model to a wider range of contexts. By combining these sources positive, negative, and easy documents. The model gains a comprehensive understanding of the global landscape, significantly enhancing its generalization capabilities and adaptability across varying query contexts. In \textbf{Phase 2}, after mastering the broader global context in \textbf{Phase 1}, the model shifts its focus to fine-tuning on the top-$a_2$ hard documents that are specific to the query. These hard documents include both positive and hard negatives documents, which are documents highly relevant to the query (e.g., identified from the top-ranked results generated by the \textbf{Phase 1} fine-tuned model) but are not labeled as positive documents. By concentrating on these challenging cases, the model hones its ability to capture intricate, query specific details and distinguish between subtle differences in complex data. This targeted refinement enhances both precision and overall performance, particularly in handling difficult or ambiguous queries. Our pipeline not only strengthens retrieval for Japanese legal texts but also outperforms on broader benchmarks, such as MS-MARCO \citep{bajaj2016ms}.\vspace{0.5em}\\
\noindent\textbf{Contributions.} Our contributions are as follows:
\begin{itemize}
\item We propose a novel text retrieval pipeline designed for Japanese legal contexts, inspired by the RepLLaMA \citep{ma2024fine}. This approach leverages large language models (LLMs) to optimize retrieval efficiency and accuracy, building on existing methods to achieve superior results. Our proposed method not only outperforms on our collected Japanese dataset but also demonstrates exceptional performance on standard benchmarks such as MS-MARCO.\\
\item We introduce a high-quality, native Japanese legal dataset. Unlike translation-based datasets, which often lose essential cultural and linguistic nuances, this dataset was developed directly in Japanese using Gemini Pro and was manually verified for accuracy. By preserving the unique characteristics of Japanese legal language, this dataset provides an essential resource for more accurate and context-aware retrieval in the Japanese legal domain.\\
\item Additionally, we extend our pipeline with an ensemble model, as illustrated in Section \ref{sec: extension}, combining various methods to further improve retrieval quality. This ensemble approach enhances the pipeline’s adaptability and robustness across different retrieval tasks, yielding the best possible results.
\item Extensive experiments have been conducted to validate the superior performance of our proposed method, demonstrating its effectiveness across diverse scenarios and languages.
\end{itemize}

\noindent
In summary, the primary contributions of this paper are the proposed retrieval pipeline tailored for Japanese legal contexts, the introduction of a carefully curated Japanese legal dataset, and an ensemble extension that further enhances retrieval performance.

\section{Preliminaries}
In this paper, we focus on advancing and optimizing traditional, long-established methods and pipelines for text retrieval tasks. By analyzing the limitations and strengths of these conventional approaches, we aim to leverage this understanding as a foundation for improvement, addressing weaknesses, and ultimately developing a more robust and efficient pipeline. There have been many approaches and methods, from traditional ad-hoc text retrieval to modern methods using embeddings, representing sentences as vectors in embedding space, developed from past decades to recent years.
\\
\\
\noindent
\textbf{Non-neural approaches}, also referred to as \textit{sparse retrieval}, These methods, decide relevance based on the frequency and occurrence of the words in the query and the documents. Non-neural approaches do not perform well in the semantic searching problem due to lexical mismatching in the answers. However, they are still useful for current SOTA models. BM25 \citep{robertson2004simple} and TF-IDF \citep{SALTON1988513} are well-known among all while BM25+\citep{robertson1994some} is currently the most effective in the approach. To overcome the lexical mismatching challenge, dense embedding is used to represent queries and documents. The main idea of this method was proposed with the LSI approach \citep{deerwester1990indexing}.
\\
\\
\noindent
\textbf{Dense Retrieval} leverages pre-trained deep neural language models to encode questions and supporting documents into continuous latent semantic vectors. This approach enables more accurate similarity matching and ranking of relevant documents. By transforming queries and documents into dense embeddings, dense retrieval facilitates efficient and precise retrieval of relevant information in response to a given query. The effectiveness of this method has been demonstrated in recent studies such as \citep{jin2023lader,sil2023primeqa}, and \citep{zhai2023contrastive}, which show that dense retrieval significantly enhances retrieval performance by improving how relevant documents are matched and ranked.
\\
\\
\noindent
\textbf{Cross-encoder approaches} utilize a backbone language model, such as BERT \citep{devlin2018bert}, RoBERTa \citep{liu2019roberta},  XLM-RoBERTa \citep{conneau2019unsupervised} or other transformer-based encoders, to capture global interactions between a query and a document \citep{ren2021pair,qu2020rocketqa}. In this method, the query and document are processed together as a single input pair, which passes through the language model to generate a joint representation that effectively captures their relationship. While the cross-encoder offers significant advantages in modeling detailed interactions between the inputs, it also has notable drawbacks, such as higher computational demands and longer training times compared to the dual-encoder approach. Despite these limitations, its ability to capture rich, global interactions makes it a highly effective technique for dense retrieval in information retrieval and related applications.
\\
\\
\noindent
\textbf{Bi-encoder approaches} The bi-encoder approach is commonly used in dense retrieval tasks, where both queries and documents are encoded into fixed-size vectors independently. Unlike cross-encoders, where a query and a document are jointly processed to capture their interactions, bi-encoders generate separate embeddings for queries and documents, making retrieval more efficient at scale. This method is particularly suitable for large datasets, as document embeddings can be precomputed, which speeds up the retrieval process during inference. The similarity between query and document embeddings is typically computed using cosine similarity or dot product. 

In the context of dense retrieval, bi-encoder models have shown promising results in various applications. For instance, in this paper \citep{reimers2019sentence} utilizes a bi-encoder architecture with BERT to efficiently handle tasks like semantic similarity and information retrieval. By encoding the query and document separately, the model is capable of quickly calculating the cosine similarity between them, making it more scalable for large-scale retrieval systems. Similarly, the paper on Vietnamese legal text retrieval \citep{pham2022multi} also adopts a bi-encoder-based transformer model to handle the complexity of legal document retrieval.

While bi-encoder models have shown great promise in dense retrieval tasks, there has been relatively little research focused on leveraging Large Language Models (LLMs) for this purpose. This presents an opportunity to harness the power of LLMs to further enhance performance. This is why we propose to integrate LLAMA into the bi-encoder pipeline in our approach, LLaMA \citep{touvron2023a, touvron2023b} is a large language model (LLM) based on a decoder-only Transformer architecture, designed for auto-regressive generation tasks. With billions of parameters and extensive pre-training on vast amounts of web data, LLaMA excels at generating predictions one token at a time. Its uni-directional nature means the model can only attend to earlier tokens in a sequence when making predictions, focusing exclusively on preceding elements. For an input sequence \( x = [t_1, t_2, \dots, t_{n-1}] \), LLaMA calculates the probability of the next token \( t_n \) by looking solely at the previous tokens. This can be expressed mathematically as \( P(t_n | t_1, \dots, t_{n-1}) \), where \( P \) represents the model’s probability of generating the next token based on all preceding tokens in the sequence. This auto-regressive approach allows the model to predict and generate coherent text step by step, maintaining a logical flow throughout the sequence. 

In addition to utilizing LLAMA, we also introduce a new training phase focused on hard negative mining. This phase aims to identify and learn from hard negatives documents that are superficially similar to relevant documents but are contextually incorrect. By incorporating this phase, our model becomes better at distinguishing between subtle differences in meaning, enhancing its ability to retrieve the most relevant documents even in cases where lexical similarity might be misleading.
\\
\\
\noindent
\textbf{Dual-encoder approaches} The dual-encoder approach involves two backbone language models, typically transformer encoder models or, more recently, Large Language Models (LLMs). One model is responsible for encoding queries, while the other encodes documents. This method maps both queries and documents into a shared vector space, where the inner product of their respective embeddings serves as an efficient similarity measure. Dual-encoders are highly scalable for large datasets due to two key mechanisms: (1) sharing weights among targets via a parametric encoder, and (2) utilizing a computationally efficient scoring function based on inner products \citep{monath2023improving,fu2023ss}. When integrating LLMs, the model can leverage the \texttt{EOS} token embedding, which acts as a contextual representation of the entire sentence. This token’s embedding encapsulates the semantic meaning of the full query or document, allowing for a richer, more accurate comparison of inputs within the shared vector space. The use of LLMs in dual-encoder architectures opens up new possibilities for improving the model's ability to capture nuanced semantic relationships.
\\
\\
\noindent
\textbf{Generative retrieval} is an emerging paradigm in text retrieval that utilizes generative models to directly produce relevant document identifiers (docids) or content for a given query GENRET \citep{sun2024learning}, DSI \citep{tay2022transformer}, DSI-QG \citep{zhuang2022bridging}. Unlike traditional retrieval methods (such as \textit{sparse} or \textit{dense retrieval}) that rely on pre-encoded document embeddings and matching them with queries via similarity measures, generative retrieval models treat the task as a sequence generation problem. These models are capable of generating document identifiers or text based on the input query by leveraging large language models (LLMs) or autoregressive language models.The key innovation of generative retrieval lies in its end-to-end nature, where the model generates a ranked list of results directly, without the need for explicit document indexing or vector search. Generative models learn to map queries to their relevant documents by generating unique docids or specific document content. This approach aims to better exploit the expressive power of modern generative models, offering the potential for improved retrieval accuracy, particularly in cases where traditional methods struggle with lexical matching or semantic understanding.
\section{Proposed Approach}
With the increasing digitization of legal text data, the NLP community has introduced numerous datasets to aid researchers in building reliable models for various legal tasks. However, some of these datasets are translations from one language to another, such as from Japanese to other languages or vice versa, which can compromise the authenticity of the original data. The sentences in these datasets are often not native but translated from foreign legal systems, leading to potential inaccuracies in reflecting the original legal reasoning and structure. These limitations highlight the need for a new large-scale, citizen-centric dataset with native content for statutory article retrieval. 

In this paper, we propose the creation of a dataset generated by Gemini 1.5 Pro, followed by manual review to ensure the highest possible level of accuracy. Additionally, building upon the insights gained from \citep{gao2021unsupervised}, we aim to extend and improve this pipeline by introducing a new phase in the training process. Specifically, we propose a fine-tuning phase, along with an additional training stage, to boost the performance of the pipeline, ensuring it achieves optimal results. These enhancements are designed to increase performance and provide significant improvements compared to the original pipeline in \citep{ma2024fine}. This approach allows the system to fully leverage the potential of Large Language Models (LLMs) for retrieval tasks and incorporates a novel phase to significantly enhance efficiency.
\subsection{Dataset Preparation}
\label{sec: data}
The dataset was created in five stages: (i) compiling a large corpus of Japanese legal articles related to employment contracts; (ii) collecting Japanese employment contracts relevant to the topics identified in stage (i); (iii) dividing the contracts into smaller chunks; (iv) matching these chunks to the corresponding legal articles in the corpus and (v) generating relevant contract chunks based on the legal articles in the corpus.\vspace{1em} \\
\textbf{(i) Law articles collection:} In the Japanese legal system, there are various levels of legal documents, including Laws, Cabinet Orders, Imperial Decrees, Ministerial Ordinances, and Rules \citep{hahn1983overview}. However, we focused solely on Laws, which represent the highest level of legal authority. This approach aligns with the classification provided in the Japanese legal framework, where Laws hold the most significant legal weight and are often prioritized in legal research \citep{hahn1983overview}. 

Laws were retrieved from the Japanese e-Gov website.\footnote{https://laws.e-gov.go.jp/}, which provides access to a variety of laws covering topics such as society, marriage, and economics. For the scope of our research, we concentrated on laws related to employment contracts. We analyzed 11 legal codes for the aforementioned topic, as presented in Table \ref{appendixx}. An ID was assigned to each set of laws based on its title, and the articles within each set were numbered sequentially to create unique article IDs. We excluded any additional regulations that were not formally enacted as of August 2024. Ultimately, we compiled a corpus \(C = \{D_1, D_2, \dots, D_n\}\), consisting of \(n\) consisting of \(n\) = 743 articles, which serve as the basic retrieval units for our study.\vspace{1em} \\
\textbf{(ii) Collection of Japanese Employment Contracts:} We gathered Japanese employment contract documents from real contracts used by Japanese companies. Initially, three different employment contract documents were collected. Using real-world contracts allows for a deeper understanding of legal choices and contractual clauses made by actual companies, providing empirical validity to our dataset \citep{sanga2014choice}. This focus on practical data helps capture nuances that may not be present in hypothetical or template-based contracts, thus making the dataset more reflective of real-life legal scenarios.
\\
\\
\newcolumntype{s}{>{\columncolor{lightgray}}p{1.2cm}} 
\begin{CJK}{UTF8}{min}
\begin{table*}[!t]
\setlength{\arrayrulewidth}{0.4mm} 
\setlength{\tabcolsep}{3pt} 
\renewcommand{\arraystretch}{1.2} 
\footnotesize 
\centering 
\caption{Example of contract section chunking}\label{exp1}
\begin{tabularx}{\columnwidth}{|s|X|} 
\hline
\textbf{Original} & 
\textbf{第7章: 退職に関する問題} \newline
{\footnotesize (Chapter 7: Retirement Issues)}\newline
\textbf{第1条: 退職期間} \newline
{\footnotesize (Article 1: Retirement Period)}\newline
\textbf{1 通常の場合、男性の定年は60歳、女性の定年は55歳です。}\newline
{\footnotesize (1. In normal cases, the retirement age is 60 years old for men and 55 years old for women.)} \newline
\textbf{2 従業員が早期退職を希望する場合、男性の定年は55歳、女性の定年は50歳です。}\newline
{\footnotesize (2. If the employee requests early retirement, the age is 55 years old for men and 50 years old for women.)} 
\\
\hline
\textbf{Chunk 1} & 
\textbf{第7章: 退職に関する問題} \newline
{\footnotesize (Chapter 7: Retirement Issues)}\newline
\textbf{第1条: 退職期間} \newline
{\footnotesize (Article 1: Retirement Period)}\newline
\textbf{1 通常の場合、男性の定年は60歳、女性の定年は55歳です。}\newline
{\footnotesize (1. In normal cases, the retirement age is 60 years old for men and 55 years old for women.)} \\
\hline
\textbf{Chunk 2} & 
\textbf{第7章: 退職に関する問題} \newline
{\footnotesize (Chapter 7: Retirement Issues)}\newline
\textbf{第1条: 退職期間} \newline
{\footnotesize (Article 1: Retirement Period)}\newline
\textbf{2 従業員が早期退職を希望する場合、男性の定年は55歳、女性の定年は50歳です。}\newline
{\footnotesize (2. If the employee requests early retirement, the age is 55 years old for men and 50 years old for women.)} \\
\hline
\end{tabularx}
\end{table*}
\end{CJK}
\textbf{(iii) Contract Chunking:} The three contracts were divided into smaller chunks based on their structure. An example of a contract section is shown in Table \ref{exp1}. This chunking process was applied to all three contracts, resulting in a total of 377 chunks. Chunking is an effective approach for handling long legal documents and has been shown to improve model accuracy in tasks like multi-label text classification, as demonstrated in prior studies \citep{chalkidis2019large}.
\\
\\
\textbf{(iv) Matching Contracts to Law Articles:} The collected chunks were annotated with plain text references to relevant law articles. Relevant rules for these chunks were identified using the Gemini 1.5 Pro to match rules with chunks. Subsequently, manual verification was conducted to remove any chunks that were not directly related to the identified rules. The combination of using a large language model (LLM) and manual verification provided a reasonable degree of accuracy for this dataset. Such a human-in-the-loop approach is recommended in legal applications of LLMs to reduce risks and improve annotation reliability, particularly in contexts requiring high precision \citep{bommasani2021opportunities}. 

Ultimately, we obtained 302 chunks, each carefully labeled with the IDs of the corresponding relevant rule articles from our corpus. The dataset was then split into training, validation, and test sets consisting of 99, 73, and 130 chunks, respectively.
\\
\\
\textbf{(v) Generating Relevant Contract Chunks:} Due to the limited amount of training data, which was insufficient for training the model to achieve optimal results, we took the additional step of automatically generating contract data from the laws. Specifically, for each law in the set, we generated five sentences that mirrored the typical content of a contract related to that law using Gemini 1.5 Pro. 

We then employed a separate prompt to re-evaluate the generated sentences to ensure they were genuinely related to the corresponding law. There were 743 laws, which initially corresponded to 3,715 contract sentences (743 × 5). However, some laws proved difficult to match with relevant contract sentences, and further re-evaluation using prompts eliminated some. As a result, we ended up with 3,172 qualified contract sentences. Generating data from language models tailored to specific domains, such as legal, has shown promise in producing high-quality, context-relevant text \citep{gururangan2020don}. 

\newcolumntype{s}{>{\columncolor{lightgray}\centering\arraybackslash}m{2.5cm}} 
\begin{CJK}{UTF8}{min}
\begin{table*}[!t]
\setlength{\arrayrulewidth}{0.4mm} 
\setlength{\tabcolsep}{5pt} 
\renewcommand{\arraystretch}{1.8} 
\footnotesize 
\centering 
\caption{Example of chapter generation for contract sentences}\label{exp2}
\begin{tabularx}{\columnwidth}{|s|X|} 
\hline
\textbf{Contract sentence} & 
\textbf{通常の場合、男性の定年は60歳、女性の定年は55歳です。} \newline
{\footnotesize (In normal cases, the retirement age is 60 years old for men and 55 years old for women.)} \\
\hline
\textbf{After generating the chapter} & 
\textbf{第7章: 退職に関する問題} \newline
{\footnotesize (Chapter 7: Retirement Issues)} \newline
\textbf{第1条: 退職期間} \newline
{\footnotesize (Article 1: Retirement Period)} \newline
\textbf{1 通常の場合、男性の定年は60歳、女性の定年は55歳です。} \newline
{\footnotesize (1. In normal cases, the retirement age is 60 years old for men and 55 years old for women.)} \\
\hline
\end{tabularx}
\end{table*}
\end{CJK}

By leveraging prompts designed for legal content, we ensured that the generated data aligns closely with the language and structure of real contracts. Next, to align with the chunking structure described in stage \textbf{(iii)}, we generated chapter data for each contract sentence, as shown in Table \ref{exp2}.

All the data generated in this step was used for training. In summary, the final dataset consists of 3,259 rows of training data, 73 rows of validation data, and 130 rows of test data.
\subsection{Proposed pipeline}
\label{sec: proposed}
In this paper, we propose a novel text retrieval pipeline that leverages large language models (LLMs) to maximize retrieval efficiency and performance. Our approach is inspired by several key advances in the field, particularly from Dense Passage Retrieval (DPR) \citep{karpukhin-etal-2020-dense}, CoCondenser \citep{gao-callan-2022-unsupervised}. Which is commonly used in dense retrievers BERT \citep{devlin2018bert}, typically use a bi-directional encoder and rely on the \texttt{<CLS>} token for text representation. However, our pipeline fully replaces Language Model (LMs) such as, BERT with a large-scale model like LLaMA, allowing us to exploit the more advanced representation capabilities of LLMs. Specifically, we adopt the architecture outlined \citep{ma2024fine}, where LLaMA, a uni-directional model, is adapted for dense retrieval by appending an end-of-sequence token \texttt{<EOS>} to input queries or documents. This allows us to compute the dense vector embeddings from the last token representation, facilitating more effective query-document similarity calculations through the dot product of their respective embeddings. The dense embedding of a query or document is computed as follows:
\[
V_T = \text{Decoder}('t_1\ t_2\ \dots\ t_k\ \texttt{<EOS>}')[-1]
\]

In this formulation, \text{Decoder}(·) represents the LLaMA model, which outputs token embeddings from the final layer for each token in the input sequence. We extract the embedding of the \texttt{<EOS>} (End-of-Sequence) token as the vector representation for the entire input sequence \(t_1, t_2, \dots, t_k\), where the sequence could represent either a query \(Q\) or a document \(D\). Given a corpus \(C = \{D_1, D_2, \dots, D_n\}\), consisting of \(n\) documents, the task is to find the most relevant document \(D\) for a given query \(Q\). The relevance between a document \(D\) and the query \(Q\) is computed by taking the dot product of their corresponding dense embeddings \(V_Q\) and \(V_D\), defined as \(\text{Sim}(Q, D) = \langle V_Q, V_D \rangle\).
\\

The model is then optimized end-to-end according to InfoNCE loss:

\begin{equation}
\begin{aligned}
\mathcal{L}(Q, D^+, \{D_N\}) &= -\log p(D = D^+ \mid Q) \\
&= -\log \frac{\exp(\text{Sim}(Q, D^+))}{\exp(\text{Sim}(Q, D^+)) + \sum\limits_{D_i^- \in \{D_N\}} \exp(\text{Sim}(Q, D_i^-))}
\label{eq:infonce_loss}
\end{aligned}
\end{equation}

Here, $D^+$ represents a document that is relevant to the query $Q$ (based on human labels), while $\{D_N\}$ denotes easy and negative documents. These $D_N$ documents consist of top-ranked documents relevant to the query $Q$ but not classified as positive. They are sampled using mechanisms such as BM25+, random selection, context matching, or even initial predictions from the LLM, and are referred to as negative documents. Additionally, there are in-batch negatives, which are documents relevant to other queries within the same training batch. These are termed easy documents and are used as part of the document dataset during the fine-tuning process in \textbf{Phase 1}. Easy documents are relatively simpler to distinguish as non-relevant. Therefore, our proposal introduces a new phase, \textbf{Phase 2}, customized to further optimize the pipeline, building on this foundation. In \textbf{Phase 2}, the $D_N$ document set shifts focus to hard negatives, which are exclusively top-ranked documents highly relevant to the query $Q$, as predicted by the LLM fine-tuned in \textbf{Phase 1}. Unlike \textbf{Phase 1}, easy documents are no longer used in the same training batch. This targeted refinement enhances the pipeline’s ability to address more challenging cases, such as hard-to-classify documents, significantly improving the model’s performance and its capability to handle complex retrieval scenarios effectively.

During the inference stage, the query is typically encoded in real-time, and the retrieval of the top-k most similar documents is performed using flat indexes from the pre-encoded corpus, leveraging techniques such as FAISS developed by Facebook Research. This process enables a rigorous evaluation of the model’s effectiveness.
\begin{figure}[!t]
    \centering
    \includegraphics[width=0.9\textwidth]{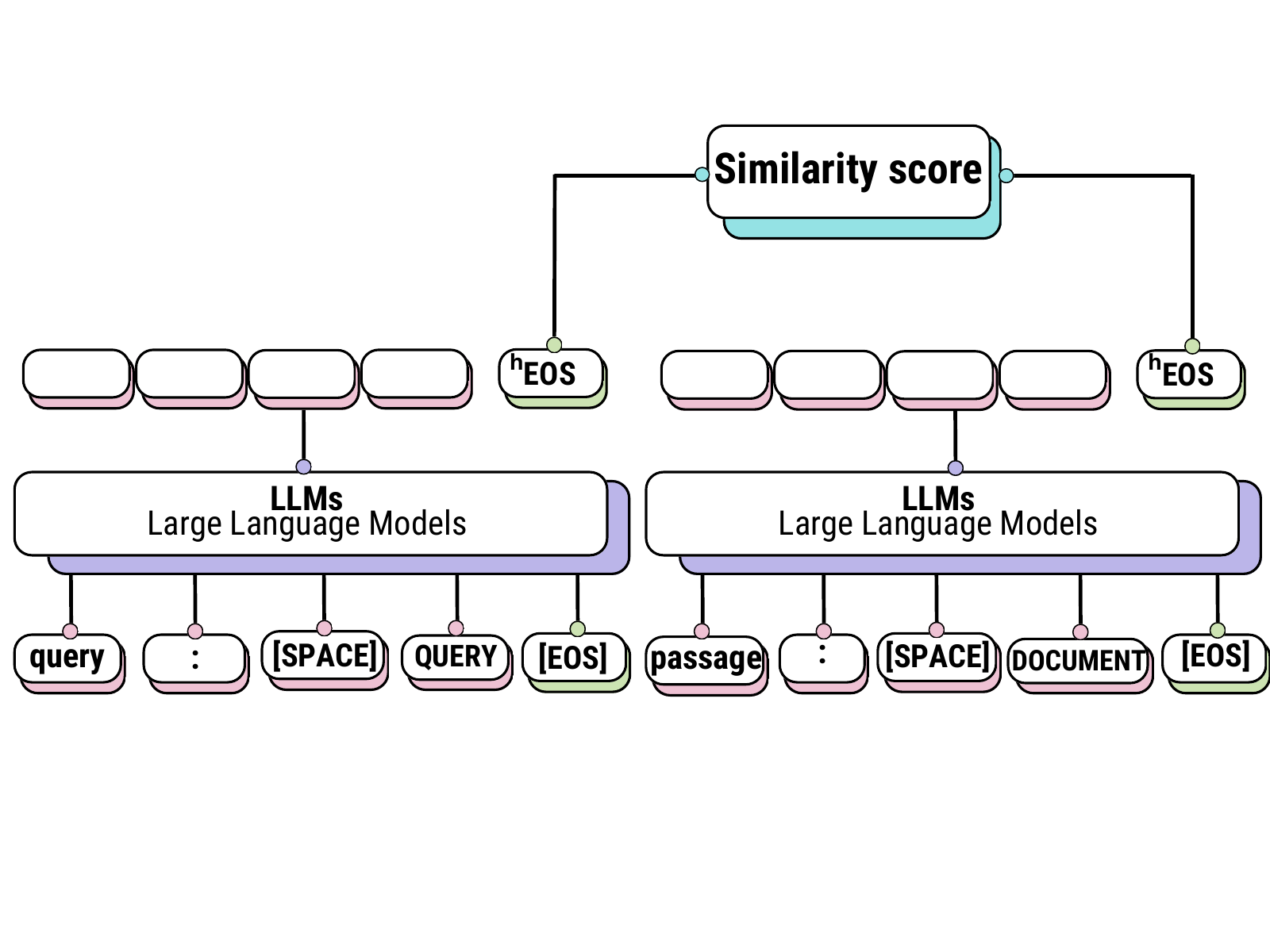} 
    \caption{Illustration of the encoding mechanism utilizing Large Language Models (LLMs). The query and document are independently transformed into dense representations through separate encoding pathways, with the hidden state at the \texttt{<EOS>} token serving as the final embedding for each. These embeddings are then employed to compute a similarity score, quantifying the semantic alignment between the query and the document.}
    \label{fig:image_2}
\end{figure}

Most retrieval models traditionally follow a three-stage framework: the first stage utilizes BM25, random selection, a context-based mechanism \citep{lu2021multi}, or even a pre-finetuned language model; the second stage
involves a language model for further processing; and the third stage typically incorporates a reranker to optimize the ranking of the top documents, often relying on another language model. This study adopts a similar three-stage approach, but with a particular focus on leveraging large language models (LLMs) within the pipeline. Special attention is given to the \textbf{Phase 2} to boost performance by training the model on more challenging and stringent datasets. The following sections provide a detailed breakdown of each module.

\subsubsection{BM25}
BM25 \citep{robertson1995okapi} is a rapid, vocabulary-based search technique widely employed by search engines like ElasticSearch. It evaluates the similarity between each document and the query to rank all documents accordingly. BM25+ \citep{robertson1994some} is an extension of the BM25 algorithm that improves retrieval effectiveness by adding a constant to the term frequency component. This modification helps prevent the relevance score from becoming zero when query terms are present infrequently, thereby enhancing performance, especially for documents with low term frequencies.

\subsubsection{Large Language Models (LLMs)}
As shown in Figure~\ref{fig:image_2} presents an encoding mechanism that employs Large Language Models (LLMs) to measure the semantic similarity between a query and a document. The query and document are independently processed through distinct encoding pathways. Each pathway begins by tokenizing the input and appending specialized tokens such as \texttt{<EOS>} to mark the end of the sequence. This configuration helps the model capture both the sequence structure and contextual boundaries. The LLMs convert these tokenized sequences into dense vector embeddings, where the hidden state corresponding to the \texttt{<EOS>} token is extracted as the final representation for each pathway. These embeddings encapsulate the semantic content of the entire input, summarizing the information from all tokens in the sequence. The final embeddings from both the query and document are then utilized to compute a similarity score. This score quantifies the semantic alignment between the two embeddings, serving as an indicator of how relevant the document is to the given query. By leveraging this approach, the system can effectively rank documents based on their contextual alignment with the query, thus enhancing the retrieval performance.

\subsubsection{In-batch negatives}
\label{sec:in_batch_negative}
Assume we have a mini-batch of \( B \) questions, each associated with a relevant document. Let \( \mathbf{Q} \) be the \( B \times d \) matrix of question embeddings, and \( \mathbf{P} \) be the \( (B \times C) \times d \) matrix of document embeddings, where \( C \) is the number of documents sampled per question. In a mini-batch of size \( B \), the similarity score matrix \( \mathbf{S} = \mathbf{Q} \mathbf{P}^T \) (with \( \mathbf{P}^T \) as the transpose of \( \mathbf{P} \)) has dimensions \( B \times (B \times C) \). Each question in the mini-batch is associated with exactly one relevant document, labeled as positive (manually by humans), while the remaining \( B \times C - 1 \) documents are considered both easy and negative. These negative documents are sampled from the top-ranking results of an existing retrieval system, whereas easy documents are derived from the top documents relevant to other queries in the mini-batch. Hence, the matrix \( \mathbf{S} \) contains the similarity scores between each question in the mini-batch and all \( B \times C \) documents, where only one document per question is positive.

\subsubsection{Global Contextualization}
\label{sec:global_contextualization}
In this phase, we assign a distinctive name that aligns with the overall process described. Specifically, we follow the methodology outlined in the DPR \citep{karpukhin-etal-2020-dense} but with a crucial modification, instead of initializing the model using BERT, we leverage a large language model, particularly LLAMA, as referenced in the RepLLaMA \citep{ma2024fine}. This adjustment is aimed at enhancing the global contextualization capacity of the model, the ability to grasp and process information from a broader, more holistic perspective, much like how humans intuitively absorb and synthesize general knowledge before diving deeper into specifics. The dataset used for training incorporates similar global knowledge as in previous methods but is augmented through the use of BM25+ and in-batch negative sampling strategies. This combination allows for more efficient contrastive learning, promoting robust representations of positive documents as well as both easy and negative documents, thereby optimizing the model for better retrieval performance. The InfoNCE loss function plays a key role in this optimization, driving the model to maximize the similarity between the query  $Q$  and the correct document  $D^+$, while minimizing its similarity to incorrect (easy, negative) documents  $D^-$, as expressed mathematically in the provided equation \ref{eq:infonce_loss}.

\subsubsection{Domain-Specific Deepening}
\label{sec: deep}
In Section \ref{sec:global_contextualization}, the model was trained using positive and a mixture of both easy and negative documents, including those derived from BM25+, as outlined in Section \ref{sec:in_batch_negative}. Easy, such as documents from different queries within the same mini-batch, enabled the model to learn generalized patterns across queries. This was computationally simpler because the model had access to data from multiple queries simultaneously (i.e., \( B > 1 \)). Mathematically, this phase is described by the similarity matrix \( \mathbf{S} \in \mathbb{R}^{B \times (B \times C)} \), where \( B \) represents the number of queries in the mini-batch, and each query is paired with \( C \) documents. In this setting, both easy and negative documents are drawn from those related to the original query itself, excluding positives, as well as from documents associated with other queries within the mini-batch, providing a broader yet simpler dataset for the model to learn from.

As we transition into this deepening phase, the model focuses on handling hard negative documents, which are significantly more challenging for the model to discriminate. These hard negatives are extracted from the top-ranking results of the model fine-tuned in \textbf{Phase 1}. In contrast to \textbf{Phase 1}, the model is now restricted to processing hard negative that are only associated with the current query, without any cross-query information from the mini-batch. Mathematically, the model now operates on a restricted similarity matrix \( \mathbf{S'} \in \mathbb{R}^{C} \), where \( B = 1 \), meaning the model only considers hard negatives from the top-ranked documents specific to the current query.

This approach forces the model to specialize in handling more difficult cases by isolating the easy data, thereby deepening its representation capabilities. The progression from easy data in \textbf{Phase 1} to hard data in this deepening phase reflects a natural learning process, where the model starts with simpler tasks and gradually moves on to more complex ones, ultimately enhancing its retrieval performance by refining its ability to distinguish subtle differences between positive and hard negative documents.

\subsection{Algorithm} 
As illustrated in Figure \ref{fig:image_1}, the top-$a_1$ documents most relevant to the query are sampled from the sparse retrieval stage. In this paper, we specifically utilize BM25+ for this purpose. The set of top-$a_1$ documents may include human-labeled positive documents as well as other relevant documents, which we designate as negative documents. During training, negatives are further enriched by incorporating easy documents, which are derived from in-batch negatives to enhance data diversity, as shown in Algorithm \ref{alg:phase1}. The resulting documents are then combined into what we refer to as the $globalDataset$, which merges positive, negative, and easy documents as described in Section \ref{sec:global_contextualization}.

\begin{algorithm}[!t]
\caption{ Global Contextualization (\textbf{Phase 1})}\label{alg:phase1}
\textbf{Input:} LLMs, Query $Q$, Corpus $C = \{D_1, D_2, \dots, D_n\}$ of size $a_0$\\[0.03cm]
\textbf{Output:} Quantized LoRA-Tuned Model $M_1$
\begin{algorithmic}
    \State \textit{\textbf{Step 1:}} Retrieve top-$a_1$ relevant documents and in-batch negatives, with $a_1 < a_0$
    \State \hspace{1em} $docsPos \gets$ $humanLabeledDos(Q, C)$
   \State \hspace{1em} $docsNegAndEasy \gets bm25\raisebox{0.2ex}{\scalebox{0.9}{+}}(Q, C) \, \cup \, in\_batch\_neg(Q, C)$
    \State \textit{\textbf{Step 2:}} Combine positives along with both negatives and easy documents.
    \State \hspace{1em} $globalDataset \gets$ $(docsPos, \hspace{0.2em} docsNegAndEasy)$
    \State \textit{\textbf{Step 3:}} Train the LLMs using LoRA and Quantization
    \State \hspace{1em} $M_1 \gets$ $qLora(LLMs, \hspace{0.2em}globalDataset)$
    \State \textbf{return} $M_1$
\end{algorithmic}
\end{algorithm}

Given the limited GPU memory available, unlike the approach in \citep{ma2024fine}, we apply quantization techniques to optimize memory usage and reduce storage requirements. Additionally, we incorporate LoRA (Low-Rank Adaptation) to enhance computational efficiency for large language models (LLMs). In this paper, we use models such as LLaMA to leverage the current capabilities of LLMs and integrate them into the retrieval pipeline.

In \textbf{Phase 1} of Algorithm \ref{alg:phase1}, we output the trained model $M_1$, which has been fine-tuned on relatively simple data. This initial phase allows the model to learn broad global contextual information from the outset, before progressing to more challenging tasks in subsequent phases.
\begin{algorithm}[!t]
\caption{Domain-Specific Deepening (\textbf{Phase 2})}\label{alg:phase2}
\textbf{Input:} Model $M_1$, Query $Q$, Corpus $C$\\[0.03cm]
\textbf{Output:} Fine-tuned Quantized LoRA Model $M_2$
\begin{algorithmic}
    \State \textit{\textbf{Step 1:}} Use $M_1$ to predict top-$a_2$ hard documents, with $a_2 < a_0$
    \State \hspace{1em} $hardNeg \gets M_1.predict(Q, C) - docsPos$
    \State \textit{\textbf{Step 2:}} Combine human-labeled positives and hard negatives
    \State \hspace{1em} $domainSpecificDataset\gets$ $(docsPos, hardNeg)$
    \State \textit{\textbf{Step 3:}} Fine-tune $M_1$ using LoRA and Quantization
    \State \hspace{1em} $M_2 \gets$ $qLora(M_1, \hspace{0.2em}domainSpecificDataset)$
    \State \textbf{return} $M_2$
\end{algorithmic}
\end{algorithm}

The model $M_1$ trained during \textbf{Phase 1} is then applied in \textbf{Phase 2} to rank the top-$a_2$ hard documents that are most relevant to the query. Unlike \textbf{Phase 1}, where relevance was based more on lexical similarity, in \textbf{Phase 2}, we focus on semantic similarity. As a result, hard negatives $(hardNeg)$ documents are no longer drawn from in-batch negatives $(in\_batch\_neg)$ which could include documents from other queries and may not be relevant. Instead, as described of Algorithm \ref{alg:phase2}, hard negatives are selected exclusively from the top-$a_2$ hard documents predicted by $M_1$, ensuring that they are closely related to the current query.

These hard negatives, combined with the positive documents, form what we call the $domainSpecificDataset$ which ensures a more focused and accurate training process. Similar to \textbf{Phase 1}, the model $M_1$ is fine-tuned in this phase using the $domainSpecificDataset$, incorporating LoRA and Quantization techniques to optimize memory and computation.

In this phase, the model is trained on more challenging and complex data, improving its ability to handle difficult cases. This targeted approach significantly boosts the performance of the retrieval pipeline, allowing it to outperform baseline methods and achieve better overall results on domain-specific, hard-to-rank datasets.
\section{Extensions}
\label{sec: extension}
In this paper, we introduce a method to significantly enhance the performance of retrieval models through the use of ensemble techniques. Our approach outperforms traditional baselines by leveraging the complementary strengths of multiple models, including fine-tuned Large Language Models (LLMs) and the BM25+ scoring method. 

The baseline retrieval approach, as illustrated in Figure \ref{fig:image_1}, follows a typical sequence: a query is matched against a corpus and a similarity function (in our case, the dot-product) is applied to compute relevance scores between the query and the documents. This dot-product operation is frequently used in retrieval tasks due to its efficiency and simplicity, and in this case, it allows for quick computation of scores based on the embeddings of both the query and the corpus. The system then ranks the documents according to these scores, generating an output list. 

However, this straightforward approach lacks the sophistication required to fully exploit the diversity of modern models. While effective, this baseline method has clear limitations. The process starts with a sparse retrieval method to process the query and corpus in the initial stage. In the first phase, a fine-tuned model, trained on the documents retrieved by BM25+, identifies the top most relevant documents, referred to as hard documents, for the second phase. These hard documents are then iteratively refined during phase two. The final output is a list of scores representing the relevance of the query to all documents in the corpus  $C$ , predicted by the model in the second phase. This approach does not allow for combining the unique strengths of different models or for learning from diverse sources of information. Thus, the final output may lack the deeper contextual understanding that could be achieved by using multiple models in tandem. To address these limitations, our paper proposes an ensemble approach, which significantly enhances performance by combining various models. 

\begin{algorithm}[!t]
\caption{Final Score using Ensemble Model}\label{alg:final_score}
\textbf{Input:}
\begin{itemize}
    \setlength{\itemindent}{0.5cm}
    \item Query $Q$, Corpus $C = \{D_1, D_2, \dots, D_n\}$ of size $a_0$
    \item The $t$ models are fine-tuned $M = \{M_1, M_2, \dots, M_t\}$
    \item Weights for each model $W = \{w_1, w_2, \dots, w_t\}$
    \item $b = \{b_1, b_2, \dots, b_{a_0}\}$ scores from BM25+ for each document
    \item Weight for BM25+ scores $\alpha$
\end{itemize}
\textbf{Output:}
\begin{itemize}
    \setlength{\itemindent}{0.5cm}
    \item Final scores $S = [s_1, s_2, \dots, s_{a_0}]$ for all documents in Corpus $C$
\end{itemize}
\vspace{0.005cm}
\begin{algorithmic}
    \State \textit{\textbf{Step 1:}} Initialize the final score array $S \gets [0, 0, \dots, 0]$ of size $a_0$
    \State \textit{\textbf{Step 2:}} For each document $D_j$ in the corpus, compute its total score
    \For{$j = 1$ to $a_0$}
        \State $s_j \gets 0$ \Comment{Initialize score for document $D_j$}
        \For{$i = 1$ to $t$}
            \State $V_{Q}^{i} \gets M_i.\text{encode}(Q)$
            \State $V_{D_j}^{i} \gets M_i.\text{encode}(D_j)$
            \State $s_j \gets s_j + w_i \times \text{Sim}(V_{Q}^{i}, V_{D_j}^{i})$
        \EndFor
        \State $s_j \gets s_j + \alpha \times b_j$ \Comment{Add BM25+ contribution for $D_j$}
        \State $S[j] \gets s_j$ \Comment{Store the score for $D_j$ in the array $S$}
    \EndFor
    \State \textbf{return} $S$
\end{algorithmic}
\end{algorithm}

Ensemble methods are widely known for their ability to improve prediction accuracy by merging the outputs of different models, each contributing its unique strengths to the final decision. This is particularly useful in the context of document retrieval, where some models might excel in understanding global context, while others focus more on query-specific details.
\begin{table*}[!t]
\centering
\caption{Comparison of different ensemble schemes: RepLLaMA\scriptsize{(phase1)} \normalsize refers to models trained only on the first phase of the pipeline. Ours\scriptsize{(ckpt*)} \normalsize represents different checkpoints from our pipeline, which has been fully fine-tuned across both phases}\label{tab:retriever_performance}
\resizebox{\textwidth}{!}{%
\begin{tabular}{llccccccc|ccc}
\toprule
\textbf{Type} & \textbf{Retriever} & \multicolumn{7}{c|}{\textbf{Recall}} & \multicolumn{3}{c}{\textbf{Other}} \\
\cmidrule(lr){3-9} \cmidrule(lr){10-12}
         &           & \textbf{R@3} & \textbf{R@5} & \textbf{R@10} & \textbf{R@20} & \textbf{R@50} & \textbf{R@100} & \textbf{R@200} & \textbf{MRR@10} & \textbf{MAP@10} & \textbf{nDCG@10}\\
\midrule
\texttt{\textbf{None}}     & Ours & 64.36 & 69.04 & 76.87 & 84.72 & 92.03 & 94.99 & 98.06 & 82.54 & 61.59 & 68.86\\
\midrule
\texttt{\textbf{Single}}   & Ours \scriptsize{(ckpt*)} & 67.69 & 74.36 & 83.42 & 87.33 & 93.32 & 95.92 & 98.21 & 83 & 64.62 & 73.35\\
\midrule
\multirow{4}{*}{\texttt{\textbf{Multi}}}  
         & RepLLaMA \scriptsize{(phase1)} \normalsize + Ours & 65.13 & 72.47 & 81.07 & 87.82 & 91.36 & 95.71 & 98.06 & 82.36 & 63.54 & 71.29\\
         & BM25\scriptsize{plus} \normalsize + RepLLaMA \scriptsize{(phase1)} \normalsize + Ours & \textbf{68.21} & \textbf{77.26} & \textbf{84.12} & \textbf{88.64} & 92.13 & 95.92 & \textbf{98.28} & 84.22 & \textbf{66.73} & \textbf{74.86}\\
         & BM25\scriptsize{plus} \normalsize + Ours \scriptsize{(ckpt*)} & 67.44 & 75.17 & 82.14 & 86.92 & \textbf{93.45} & \textbf{96.37} & 98.24 & \textbf{86.81} & 65.86 & 74.38\\
\bottomrule
\end{tabular}%
}
\end{table*}

In our ensemble framework, we combine BM25+ scores with embeddings produced by fine-tuned LLMs. This allows us to integrate both traditional frequency-based retrieval (BM25+) and the rich semantic embeddings from LLMs, leading to a more comprehensive and nuanced understanding of document relevance. Our approach improves upon the baseline by taking advantage of the synergy between these methods.

We evaluate the performance of our proposed ensemble retrieval approach using a customized Legal Japanese dataset. Our experiments demonstrate the advantages of combining various retrieval strategies to improve recall, precision, and ranking metrics across different evaluation levels. Table \ref{tab:retriever_performance} provides a comprehensive comparison of retrieval performance across several ensemble configurations, highlighting the effectiveness of our approach.
\\
\\
\texttt{\textbf{None}} This represents the baseline performance without any ensemble, where a single model is fine-tuned through two phases using LoRa-based quantization on LLaMA-2. Due to resource constraints, we apply 16-bit quantization on LLaMA-2 and incorporate the LoRa mechanism to enhance computational efficiency and reduce GPU memory usage.This approach delivers competitive results, with a Recall@10 of 76.87 and an MRR@10 of 82.54. Other metrics, such as Recall@3 at 64.36 and Recall@5 at 69.04, indicate reasonable short-range recall, while Recall@200 reaches 98.06, demonstrating strong overall coverage. Despite its decent performance across most metrics, the system's dependency on a single similarity function constrains its effectiveness in diverse retrieval contexts. This limitation becomes evident when compared to more advanced ensemble methods, particularly in achieving higher MAP@10 (61.59) and nDCG@10 (68.86).
\\
\\
\texttt{\textbf{Single}} This represents a simple ensemble strategy where we combine predictions from different checkpoints of a model fine-tuned across two phases. In this setup, we combine two distinct checkpoints, each of which excels in different recall metrics (R@k). For instance, checkpoint 1 performs better for smaller values of k, meaning it excels at predicting top candidates, while checkpoint 2 performs better for larger k values, indicating it is more accurate when predicting broader candidate sets. By assigning weighted scores to each checkpoint, based on our experiments, we observe that the combination of checkpoint 1 and checkpoint 2 (with weights $w_1$ = 0.6 and $w_2$ = 0.4) yields the best performance across multiple evaluation metrics. This weighted ensemble achieves the most balanced results for recall at different levels of k, as discussed in our algorithm \ref{alg:final_score}. Recall@10 increases to 83.42, while Recall@3 and Recall@5 also see gains at 67.69 and 74.36, respectively. The MRR@10 improves to 83.00, with MAP@10 and nDCG@10 jumping to 64.62 and 73.35, highlighting the added precision and ranking quality of this method. This approach effectively leverages variations in fine-tuning across checkpoints, creating complementary models that together yield a more robust retrieval framework.
\\
\\
\texttt{\textbf{Multi}} This ensemble category explores more complex combinations of models and retrieval schemes:
\begin{itemize}
    \item RepLLaMA \scriptsize{(phase1)} \normalsize + Ours: This setup pairs a model trained only on the first phase with our fully fine-tuned model. It emphasizes lexical recall and global context, achieving a Recall@10 of 81.07 and an MRR@10 of 82.36. While it performs well, particularly at Recall@20 (87.82), it falls short compared to ensembles incorporating BM25+ in metrics like MAP@10 (63.54) and nDCG@10 (71.29).
    \item BM25\scriptsize{plus} \normalsize + RepLLaMA\scriptsize{(phase1)} \normalsize + Ours: Introducing BM25+ enhances the ensemble by integrating a traditional lexical retrieval component, which complements the semantic strengths of the large language models. This combination results in the highest Recall@10 at 84.12 and a peak MRR@10 of 84.22. Early recall metrics, such as Recall@3 and Recall@5, also improve significantly, reaching 68.21 and 77.26. The MAP@10 of 66.73 and nDCG@10 of 74.86 further validate the ensemble's superior ranking performance.
    \item BM25\scriptsize{plus} \normalsize + Ours\scriptsize{(ckpt*)}\normalsize: This configuration similarly benefits from the addition of BM25+, achieving a strong Recall@10 of 82.14 and an impressive Recall@100 of 96.37. The MRR@10 stands at 86.81, indicating excellent precision. The MAP@10 (65.86) and nDCG@10 (74.38) metrics showcase its balanced retrieval effectiveness, underscoring the advantages of blending lexical and semantic models.
    \item Ours\scriptsize{(llama2)} \normalsize + Ours\scriptsize{(llama3)}\normalsize: Combination of pretrained models (LLaMA-2 and LLaMA-3), although not implemented in this study due to hardware and resource constraints, combining outputs from models like LLaMA-2 and LLaMA-3 represents a promising future direction. LLaMA-3, with its enhanced semantic understanding, is expected to complement LLaMA-2 by bringing additional diversity to the ensemble. Despite the lack of experimental results for this configuration, its inclusion in our pipeline design highlights the flexibility and scalability of our approach.
\end{itemize}

The results in Table \ref{tab:retriever_performance} underline the importance of leveraging diverse retrieval strategies. The integration of BM25+ with fine-tuned LLMs consistently delivers superior performance by balancing lexical and semantic contexts. Moreover, our weighted checkpoint ensemble demonstrates that even within a single model's training pipeline, there are opportunities to enhance retrieval effectiveness through selective combination.

While our current study achieves state-of-the-art results, the absence of an implementation for Ours\scriptsize{(llama2)} \normalsize + Ours\scriptsize{(llama3)} \normalsize due to resource constraints presents a limitation. However, this configuration remains an exciting avenue for future research, particularly as hardware capabilities advance and cost barriers decrease. Combining outputs from multiple LLM generations could further improve recall and ranking metrics, especially in datasets requiring deep contextual understanding.

Overall, our experiments validate the effectiveness of ensemble approaches in text retrieval tasks. By incorporating complementary models and techniques, such as BM25+ and fine-tuned LLMs, we achieve significant improvements across key metrics. The modularity of our framework ensures adaptability to future advancements, including the integration of more sophisticated LLMs like LLaMA-3. This flexibility positions our method as a robust and scalable solution for retrieval tasks in complex domains such as legal document processing.

\section{Experiments}
We conducted experiments on both our customized Legal Japanese dataset, introduced in Section \ref{sec: data}, and split of the MS MARCO passage dataset, to assess the effectiveness of our proposed multi-stage text retrieval pipeline built around RepLLaMA as detailed in this paper \citep{ma2024fine}.
\subsection{Legal Japanese Retrieval}
\label{legaljp}
\textbf{Dataset} We trained our retriever using the LLaMA-2 model on our customized Legal Japanese dataset, which contains approximately 3.5k training examples. As highlighted in Section \ref{sec: proposed}, the inclusion of both negatives and easy documents from both BM25+ results and in-batch negatives is a critical element in the training process. These negatives and easy documents ensure greater diversity in the training set, contributing to a more robust model. The model’s performance was evaluated on a test set of 130 queries from the Legal Japanese dataset, with Recall, MRR@10, MAP@10 and nDCG@10 being the key evaluation metrics.
\\
\\
\textbf{Implementation Details} \label{impl} We initialized our models with the LLaMA-2-7B checkpoint and conducted the training on 2×24GB RTX 4090 GPUs. The final layer’s representation of the \texttt{<EOS>} token was extracted as the dense representation, with a dimensionality of 4096. During both training and inference, these dense representations were normalized into unit vectors, ensuring an L2-norm of 1. After pre-encoding the entire corpus, we employed a flat index for brute-force search, similar to the method used in RepLLaMA.
One of the main challenges in fine-tuning large language models (LLMs) for retrieval is the high GPU memory cost associated with contrastive learning, particularly due to the need for large batch sizes when incorporating in-batch negatives. To mitigate this, we utilized several memory-efficient techniques such as LoRA \citep{hu2021lora}, Flash Attention \citep{dao2023flashattention}, and gradient checkpointing, which significantly reduced memory usage. Additionally, we applied 16-bit quantization to LLaMA for both phases of the retriever training process.

For \textbf{Phase 1}, we sampled 50 relevant documents from BM25+ results, and during in-batch training, we selected 30 documents from the top-50 results for each query. This resulted in 30×$B$ documents per query, where $B$ is the batch size. In \textbf{Phase 1}, we set $B$=2 due to resource limitations. After completing \textbf{Phase 1}, \textbf{Phase 2} began with sampling the top 50 relevant documents from the model fine-tuned in \textbf{Phase 1}. However, unlike \textbf{Phase 1}, \textbf{Phase 2} used a batch size of $B$=1, not due to resource constraints, but as mentioned in Section \ref{sec: deep}, this setup was necessary to focus the model on learning more challenging contexts and complex data. In this phase, the hard negative documents were drawn from documents directly related to the query rather than from irrelevant documents, making the training more stringent and enhancing the model’s effectiveness.

To further optimize training, we employed gradient accumulation steps, which helped mitigate the limitations of using a small batch size ($B$=1), ensuring that the model still benefited from more substantial learning steps despite resource constraints.
\begin{table*}[!t]
\centering
\caption{Demonstrating the effectiveness of our proposed method for leveraging LLMs in text retrieval using QLoRA on a customized Japanese Legal Corpus, a comparative analysis showing superior performance over existing methods.}\label{tab: ja_eval}
\resizebox{\textwidth}{!}{
\begin{tabular}{lccccccc|ccc}
\toprule
\multirow{2}{*}{\textbf{Method}} & \multicolumn{7}{c|}{\textbf{Recall}} & \multicolumn{3}{c}{\textbf{Other}} \\
\cmidrule(lr){2-8} \cmidrule(lr){9-11}
& \textbf{R@3} & \textbf{R@5} & \textbf{R@10} & \textbf{R@20} & \textbf{R@50} & \textbf{R@100} & \textbf{R@200} & \textbf{MRR@10} & \textbf{MAP@10} & \textbf{nDCG@10} \\
\midrule
\multicolumn{11}{l}{\textit{Sparse retrieval}} \\
TF-IDF \citep{SALTON1988513} & 42.82 & 46.18 & 55.11 & 63.19 & 73.29 & 78.07 & 83.76 & 72.4 & 39.17 & 46.52 \\
BM25+ \citep{robertson1994some} & 37.82 & 41.09 & 50.59 & 57.10 & 65.84 & 73.51 & 81.45 & 58.11 & 29.76 & 38.29 \\
\midrule
\multicolumn{11}{l}{\textit{Dense retrieval}} \\
Siamese \citep{reimers-gurevych-2019-sentence} & 46.79 & 53.56 & 62.16 & 67.23 & 80.66 & 86.64 & 92.08 & 53.64 & 39.61 & 49.37 \\
Two-Towers Siamese \citep{yang-etal-2020-multilingual} & 47.82 & 55.63 & 62.23 & 70.12 & 80.29 & 85.74 & 91.99 & 57.74 & 41.76 & 51.28 \\
GC-DPR \citep{karpukhin-etal-2020-dense} & 42.56 & 46.42 & 53.05 & 63.10 & 76.18 & 82.49 & 87.96 & 75.98 & 37.47 & 46.38 \\
SXLM-RoBERTa \citep{pham2022multi} & 65.38 & 68.33 & 75.91 & 78.93 & 87.99 & 92.36 & 96.71 & 77.39 & 62.15 & 68.69 \\
CoCondenser \citep{gao-callan-2022-unsupervised} & 60.77 & 62.02 & 70.89 & 77.10 & 82.26 & 87.34 & 92.92 & 73.44 & 57.11 & 65.01 \\
3SRM \citep{sasazawa2023text} & 39.87 & 42.38 & 50.13 & 59.95 & 76.85 & 86.76 & 90.08 & 48.16 & 34.88 & 42.15 \\
RepLLaMA \citep{ma2024fine} & 60.26 & 66.86 & 73.73 & 81.78 & 89.96 & 94.18 & 97.81 & 81.58 & 59.02 & 66.38 \\
\textbf{Ours} & 64.36 & \textbf{69.04} & \textbf{76.87} & \textbf{84.72} & \textbf{92.03} & \textbf{94.99} & \textbf{98.06} & \textbf{82.54} & 61.59 & \textbf{68.86} \\
\midrule
\multicolumn{11}{l}{\textit{Generative retrieval}} \\
DSI \citep{tay2022transformer} & \textbf{66.03} & 68.86 & 75.21 & 80.34 & 86.38 & 89.19 & 92.90 & 68.16 & \textbf{66.06} & 65.39 \\
DSI-QG \citep{zhuang2022bridging} & 15.75 & 23.29 & 28.77 & 39.49 & 56.85 & 68.72 & 84.25 & 13.64 & 13.63 & 16.75 \\
\bottomrule
\end{tabular}
}
\end{table*}
In addition, we experimented with an ensemble model pipeline, as discussed in Section \ref{sec: extension}, which demonstrated the effectiveness of combining multiple models. By aggregating their predictions, we achieved improved performance across the Japanese legal dataset, showing the benefits of model ensembling in increasing both precision and recall.
\\
\\
\noindent \textbf{In-Domain Evaluation} Table \ref{tab: ja_eval} highlights the superior performance of our proposed method compared to sparse, dense, and generative retrieval approaches on the customized Japanese legal corpus. Sparse retrieval methods, such as TF-IDF \citep{SALTON1988513} and BM25+ \citep{robertson1994some}, rely solely on keyword frequency and fail to capture semantic nuances, leading to poor results across all metrics. For example, at Recall@3, our method scores 64.36, outperforming BM25+ by 26.54 points and TF-IDF by 21.54 points. This advantage persists at Recall@200, where our method achieves 98.06, surpassing BM25+ by 16.61 points and TF-IDF by 14.30 points.

Dense retrieval methods, including Siamese \citep{reimers-gurevych-2019-sentence}, Two-Tower Siamese \citep{yang-etal-2020-multilingual}, GC-DPR \citep{karpukhin-etal-2020-dense}, SXLM-RoBERTa \citep{pham2022multi}, CoCondenser \citep{gao-callan-2022-unsupervised}, and 3SRM \citep{sasazawa2023text} refers to the Three-Stage Re-Ranking Model, rely on language models (LMs) to create semantic embeddings but are consistently outperformed by our method due to its incorporation of \textbf{Phase 2} training with (LLMs). This additional training phase improves the pipeline by enabling better handling of complex queries and ambiguous document relationships. Notably, RepLLaMA, which uses large language models (LLMs) rather than standard LMs, is one of the strongest baselines but is also outperformed at every recall level. At Recall@3, our method surpasses RepLLaMA by 4.10 points (64.36 vs. 60.26), and at Recall@200, we maintain a lead of 0.25 points (98.06 vs. 97.81).

At Recall@20, our method achieves 84.72, surpassing GC-DPR by 21.62 points, CoCondenser by 7.62 points, Two-Tower Siamese by 14.60 points, and 3SRM by 24.77 points. While SXLM-RoBERTa performs slightly better at Recall@3 (65.38 vs. 64.36) and MAP@10 (62.15 vs. 61.59) due to its focus on challenging datasets with XLM-RoBERTa-base, our method outperforms it at higher recall levels, such as Recall@200 (98.06 vs. 96.71), thanks to the added semantic depth provided by (LLMs).

Generative retrieval methods, such as DSI, demonstrate strong performance at lower recall levels, achieving 66.03 at Recall@3, slightly surpassing our proposed method (64.36). However, their effectiveness diminishes as recall requirements increase. Notably, our method outperforms DSI by 5.16 points at Recall@200 (98.06 vs. 92.90), showcasing its superior scalability for higher-recall tasks.

Additionally, the generative retrieval method DSI-QG \citep{zhuang2022bridging} attempts to address the limitations of DSI by tackling the issue of documents without associated queries in the training set. In this approach, the authors generate new queries for individual documents and group them under a symbolic docid representing their shared topic or thematic connection. While this concept is innovative, it encounters significant challenges in the context of Japanese datasets. Specifically, the reliance on docT5query \citep{nogueira2019doc2query}, a baselines, to generate queries introduces a critical bottleneck. Generating high-quality queries for Japanese is hindered by the linguistic complexity of the language and the limited semantic understanding of the baselines. This results in suboptimal training data and severely impacts the method’s effectiveness. As illustrated in Table \ref{tab: ja_eval}, DSI-QG consistently underperforms compared to all other methods across every metric.

In contrast, our approach sets the benchmark for ranking metrics, achieving state-of-the-art performance with MRR@10 (82.54) and nDCG@10 (68.86), outperforming all baselines. These results underscore the robustness of our proposed pipeline, particularly highlighting the pivotal role of \textbf{Phase 2} training, which integrates QLoRA-based fine-tuning and the LLaMA2 model. This combination significantly enhances the system’s ability to handle complex queries, model intricate semantic relationships, and deliver superior results across both recall and ranking metrics.

\subsection{Benchmark Datasets Retrieval} 
\textbf{Dataset} Given our limited computational resources, we conducted retrieval experiments on split of the MS MARCO dataset \citep{DBLP:journals/corr/NguyenRSGTMD16}, which includes around 1.01 million rows, with a specific subset of 102k rows allocated for training, validation, and testing. For our purposes, we used a split of 17,132 rows, divided into 15,270 for training, 1,000 for validation, and 862 for testing. The corpus consisted of approximately 134,000 documents, from which we sampled regular negatives in phase one and more challenging hard negatives in phase two, as detailed in Section \ref{impl}.
\\
\\
\textbf{Implementation Details} Model training employed LLaMA-2-7B checkpoints on two NVIDIA RTX 4090 GPUs (24GB each), optimized with Quantization and LoRA techniques to fit our constraints. Unlike our prior work with Japanese legal datasets, which required a \textit{max\_seq\_length} of 512 due to longer query and document lengths, we adjusted \textit{max\_seq\_length} to 128 for the MS MARCO dataset, reflecting the shorter average text length. This adjustment enabled us to configure a batch size ($B$) of 8 during \textbf{Phase 1}, with each query associated with a total of 64 documents. This consisted of 8 top-ranked relevant documents (from the top 15 sampled using BM25+) for the target query and 56 additional documents (7 $\times$ $B$) drawn from other queries within the same mini-batch. This configuration balanced relevance and variety, enriching the model’s exposure to both relevant and non-relevant examples.

In \textbf{Phase 2}, we refined the training approach by setting the batch size to $B = 1$ to enable the model to focus on high-relevance, hard negative documents. Each query in this phase was paired with a selection of 10 documents, including both positive and challenging negatives, sampled from the top 15 results identified by the \textbf{Phase 1} trained model. This rigorous selection allowed the model to concentrate on nuanced relevance distinctions, enhancing its ability to prioritize the most challenging retrieval cases and progressively improving retrieval accuracy across increasingly difficult documents.
\begin{table*}[!t]
\centering
\caption{Illustrating the efficiency of our proposed approach for utilizing LLMs in retrieval systems, we employed QLoRA on the training split of the MS-MARCO passage dataset and evaluated its performance relative to current techniques.}\label{tab: benchmark_eval}
\resizebox{\textwidth}{!}{
\begin{tabular}{lccccccc|ccc}
\toprule
\multirow{2}{*}{\textbf{Method}} & \multicolumn{7}{c|}{\textbf{Recall}} & \multicolumn{3}{c}{\textbf{Other}} \\
\cmidrule(lr){2-8} \cmidrule(lr){9-11}
& \textbf{R@3} & \textbf{R@5} & \textbf{R@10} & \textbf{R@20} & \textbf{R@50} & \textbf{R@100} & \textbf{R@200} & \textbf{MRR@10} & \textbf{MAP@10} & \textbf{nDCG@10} \\
\midrule
\multicolumn{11}{l}{\textit{Sparse retrieval}} \\
TF-IDF \citep{SALTON1988513} & 26.97 & 37.14 & 52.82 & 62.05 & 74.98 & 83.58 & 90.35 & 43.76 & 23.22 & 30.55 \\
BM25+ \citep{robertson1994some} & 28.96 & 39.40 & 49.96 & 57.87 & 68.14 & 73.51 & 78.08 & 49.34 & 24.52 & 30.95 \\
\midrule
\multicolumn{11}{l}{\textit{Dense retrieval}} \\
Siamese \citep{reimers-gurevych-2019-sentence} & 39.83 & 53.36 & 72.25 & 82.59 & 92.23 & 95.71 & 97.22 & 35.03 & 33.95 & 43.44 \\
Two-Towers Siamese \citep{yang-etal-2020-multilingual} & 39.13 & 54.41 & 71.43 & 82.73 & 91.98 & 95.94 & 97.79 & 34.87 & 34.04 & 43.28 \\
GC-DPR \citep{karpukhin-etal-2020-dense} & 30.76 & 40.45 & 51.62 & 61.71 & 72.66 & 78.96 & 86.62 & 49.16 & 25.32 & 31.95 \\
SXLM-RoBERTa \citep{pham2022multi} & 34.51 & 45.39 & 57.21 & 66.96 & 78.79 & 84.09 & 88.88 & 30.15 & 28.99 & 36.14 \\
CoCondenser \citep{gao-callan-2022-unsupervised} & 29.23 & 34.85 & 45.85 & 54.85 & 66.74 & 73.51 & 81.99 & 25.95 & 24.57 & 30.02 \\
3SRM \citep{sasazawa2023text} & 43.08 & 55.52 & 71.63 & 81.42 & 89.73 & 93.56 & 95.19 & 36.73 & 35.64 & 44.64 \\
RepLLaMA \citep{ma2024fine} & 47.82 & 62.96 & 77.55 & 87.51 & 94.72 & 97.16 & 98.67 & 55.13 & 42.16 & 51.09 \\
\textbf{Ours} & \textbf{50.21} & \textbf{66.13} & \textbf{80.71} & \textbf{89.54} & \textbf{95} & \textbf{97.87} & \textbf{99.25} & \textbf{55.15} & \textbf{43.97} & \textbf{53.22} \\
\midrule
\multicolumn{11}{l}{\textit{Generative retrieval}} \\
DSI \citep{tay2022transformer} & 14.25 & 17.59 & 23.47 & 28.62 & 35.87 & 40.85 & 46.58 & 13.71 & 13.61 & 15.54 \\
DSI-QG \citep{zhuang2022bridging} & 39.62 & 53.55 & 70.77 & 78.85 & 86.91 & 91.44 & 94.26 & 34.93 & 34.34 & 43.30 \\
\bottomrule
\end{tabular}
}
\end{table*}
\\
\\
\textbf{In-Domain Evaluation} Table \ref{tab: benchmark_eval} highlights the effectiveness of our proposed retrieval approach, which consistently outperforms existing methods  across the training split of the MS MARCO passage dataset. 

Sparse retrieval techniques like TF-IDF and BM25+ struggle significantly, achieving low recall and ranking scores due to their inability to capture semantic nuances. In comparison, our method demonstrates a clear advantage, particularly at lower recall levels. For example, while BM25+ achieves Recall@10 of 49.96, our model achieves 80.71, marking a substantial improvement. Dense retrieval models, such as Siamese and Two-Towers Siamese, perform better than sparse methods but still lag behind our approach, particularly in ranking metrics like MRR@10, where our model achieves 55.15, indicating a notable enhancement in ranking relevance. 

Among recent advancements, RepLLaMA shows strong performance at higher recall levels, coming close to ours at Recall@100 and Recall@200. However, at lower thresholds like Recall@10 and in ranking metrics like nDCG@10, our method maintains an edge, highlighting its ability to retrieve and rank relevant documents with precision in diverse scenarios. 

Generative retrieval methods, such as DSI \citep{tay2022transformer}, exhibit particularly low performance, even falling behind traditional methods like TF-IDF and BM25+. This striking result highlights a fundamental limitation of generative approaches for retrieval tasks on MS-MARCO. One possible explanation is the way we split the dataset: the test set may contain documents related to queries that were unseen or only sparsely represented in the training set. Such splits likely exacerbate the limitations of DSI, as it relies heavily on having seen similar documents during training. This finding underscores a critical shortcoming of the generative retrieval paradigm, which performs well in scenarios where the test set closely resembles the training data but struggles significantly when tasked with retrieving unseen or underrepresented documents. Consequently, DSI’s inability to generalize effectively to out-of-distribution test cases reinforces its reliance on data overlap between training and testing phases.

In contrast to the experiments on the Japanese dataset in Section \ref{legaljp}, where the pretrained model struggled with query generation due to linguistic complexities, the use of docT5query \citep{nogueira2019doc2query} in the MS-MARCO setting demonstrates its strength. Here, the model generates more suitable and semantically aligned queries for documents, leading to improved downstream results compared to DSI. However, even with this improvement, our proposed method consistently outperforms DSI-QG \citep{zhuang2022bridging} across all metrics, as demonstrated in Table \ref{tab: benchmark_eval}. These findings highlight the robustness and adaptability of our approach, proving its effectiveness not only in handling challenging queries but also in surpassing the generative retrieval paradigm in diverse experimental settings.

Notably, our approach not only excels in tasks focused on Japanese-language datasets, where it was initially tested, but also demonstrates outstanding generalization by outperforming strong baselines on the MS-MARCO benchmark. These results showcase the adaptability and robustness of our retrieval framework, demonstrating superior performance not only in challenging high-recall tasks but also in ranking-focused metrics, thanks to the effective integration of QLoRA fine-tuning and LLaMA-2, solidifying its position as a state-of-the-art approach for retrieval on the split MS-MARCO benchmark.
\section{Conclusion}

This paper proposes a novel method for Japanese legal text retrieval, combining the introduction of a specialized dataset for this domain with a tailored two-phase pipeline. In the first phase, the model is trained to develop a broad understanding of global contexts, enhancing its generalization and adaptability to diverse queries. In the second phase, the model is fine-tuned to effectively handle complex, domain-specific legal queries. Empirical comparisons against various baselines demonstrate the superior performance of the proposed method. In future work, several avenues will be explored. First, we will experiment with incorporating state-of-the-art large language models (LLMs) such as LLaMA3 into this pipeline and evaluate their performance in ensemble approaches, as discussed in Section \ref{sec: extension}. Second, while this study focuses on generating diverse datasets with easy, negatives and hard negatives documents, future efforts will prioritize increasing the inclusion of positive documents by identifying and leveraging those most relevant to the given queries. These positives, generated using LLMs, will further enhance the training data diversity and improve model performance. Finally, we plan to integrate an advanced reranker, such as RankLLaMA, into the pipeline to optimize retrieval metrics, particularly Recall@k for low k values.

\bibliography{sn-bibliography}

\begin{appendices}
\section{Legal Codes}
Table \ref{appendixx} overview of the 11 employment-related legal codes from the Japanese e-Gov database, covering key areas such as wage payments, retirement, and labor standards. Each code was assigned a unique ID, with articles sequentially numbered to form a 743-article corpus for analysis.
\newcolumntype{s}{>{\columncolor{lightgray}}p{1.5cm}}

\begin{CJK}{UTF8}{min}
\begin{table}[!t]
\setlength{\arrayrulewidth}{0.4mm} 
\setlength{\tabcolsep}{4pt} 
\renewcommand{\arraystretch}{1.4} 
\footnotesize 
\centering
\caption{11 Code of Corpus}\label{appendixx}
\begin{tabularx}{\textwidth}{|s|X|c|}
\hline
\textbf{No.} & \textbf{Japanese Law Name} & \textbf{Number of Laws} \\
\hline
1 & 賃金の支払の確保等に関する法律 \newline
{\footnotesize (Law on Securing Wage Payments)} & 21 \\
\hline
2 & 中小企業退職金共済法 \newline
{\footnotesize (Small and Medium-Sized Enterprise Retirement Allowance Law)} & 104 \\
\hline
3 & 育児休業、介護休業等育児又は家族介護を行う労働者の福祉に関する法律 \newline
{\footnotesize (Law on Welfare for Workers Taking Childcare or Family Care Leave)} & 77 \\
\hline
4 & 家内労働法 \newline
{\footnotesize (Home Work Law)} & 35 \\
\hline
5 & 労働時間等の設定の改善に関する特別措置法 \newline
{\footnotesize (Special Measures Law for Improving Working Hours Settings)} & 17 \\
\hline
6 & 短時間労働者及び有期雇用労働者の雇用管理の改善等に関する法律 \newline
{\footnotesize (Law on Improving Employment Management of Part-Time and Fixed-Term Workers)} & 32 \\
\hline
7 & 労働契約法 \newline
{\footnotesize (Labor Contract Law)} & 21 \\
\hline
8 & 労働基準法 \newline
{\footnotesize (Labor Standards Law)} & 124 \\
\hline
9 & 労働保険の保険料の徴収等に関する法律 \newline
{\footnotesize (Law on Collection of Labor Insurance Premiums)} & 54 \\
\hline
10 & 労働者災害補償保険法 \newline
{\footnotesize (Workers’ Accident Compensation Insurance Law)} & 98 \\
\hline
11 & 職業安定法 \newline
{\footnotesize (Employment Security Law)} & 118 \\
\hline
\end{tabularx}
\end{table}

\end{CJK}
\section{Evaluation Metrics}
To comprehensively evaluate the performance of our multi-stage document retrieval framework, we employ a set of widely adopted metrics in information retrieval. These metrics capture various aspects of retrieval quality, including recall, ranking accuracy, and relevance distribution. We report performance using the following metrics:
\section*{1. Recall (R@k)}
\noindent\textbf{Definition:} Recall at rank $k$, denoted as R@k, is the proportion of relevant documents retrieved within the top $k$ results. It is calculated as:
\[
R@k = \frac{\text{Number of relevant documents retrieved in the top } k}{\text{Total number of relevant documents}}
\]
\noindent\textbf{Explanation:} Recall focuses on the system's ability to retrieve as many relevant documents as possible. By reporting R@3, R@5, R@10, R@20, R@50, R@100, and R@200, we evaluate the effectiveness of the retrieval system at various depths, providing insights into both early and deeper retrieval performance.

\vspace{0.5cm}

\section*{2. Mean Reciprocal Rank (MRR@k)}
\noindent\textbf{Definition:} The Mean Reciprocal Rank at rank $k$, MRR@k, is the average of the reciprocal ranks of the first relevant document retrieved for each query, limited to the top $k$ results. It is defined as:
\[
MRR@k = \frac{1}{|Q|} \sum_{i=1}^{|Q|} \frac{1}{\text{rank}_i}
\]
where $|Q|$ is the number of queries, and $\text{rank}_i$ is the position of the first relevant document for query $i$ in the top $k$ results.

\noindent\textbf{Explanation:} MRR emphasizes the ranking position of the first relevant document. By using MRR@10, we focus on the system's ability to present a relevant document as early as possible within the top 10 results. Higher MRR values indicate that relevant documents are ranked closer to the top.

\vspace{0.5cm}

\section*{3. Mean Average Precision (MAP@k)}
\noindent\textbf{Definition:} MAP at rank $k$, MAP@k, is the mean of the average precision scores calculated for each query, where average precision (AP) is defined as:
\[
AP = \frac{\sum_{j=1}^k \text{Precision@j} \cdot \text{rel}(j)}{\text{Total number of relevant documents}}
\]
Here, $\text{Precision@j}$ is the precision at rank $j$, and $\text{rel}(j)$ is an indicator function that is 1 if the document at rank $j$ is relevant and 0 otherwise. MAP@k is then:
\[
MAP@k = \frac{1}{|Q|} \sum_{i=1}^{|Q|} AP_i
\]
\noindent\textbf{Explanation:} MAP provides a single-figure measure of quality across recall levels. By focusing on MAP@10, we measure the average precision of the top 10 results, giving insight into both the ordering of relevant documents and the precision of the retrieval system.

\vspace{0.5cm}

\section*{4. Normalized Discounted Cumulative Gain (nDCG@k)}
\noindent\textbf{Definition:} nDCG@k is a ranking metric that considers the graded relevance of documents, with a higher weight assigned to more relevant documents appearing higher in the ranking. It is calculated as:
\[
nDCG@k = \frac{DCG@k}{IDCG@k}
\]
where Discounted Cumulative Gain (DCG) at rank $k$ is:
\[
DCG@k = \sum_{j=1}^k \frac{2^{\text{rel}(j)} - 1}{\log_2(j + 1)}
\]
and Ideal DCG (IDCG@k) is the DCG obtained by an ideal ranking of documents. The function $\text{rel}(j)$ denotes the graded relevance of the document at rank $j$.

\noindent\textbf{Explanation:} nDCG@10 evaluates the quality of the ranking, particularly emphasizing the placement of highly relevant documents at the top of the ranked list. This metric is essential when relevance levels are not binary, allowing for a nuanced assessment of ranking performance.
\\
\\
These metrics collectively capture different dimensions of retrieval performance, from the ability to retrieve all relevant documents (Recall) to prioritizing relevant documents early (MRR, MAP) and handling graded relevance (nDCG). This comprehensive evaluation framework ensures that our system's performance can be thoroughly analyzed and compared to existing approaches.

\end{appendices}

\end{document}